\documentclass[ aps,  prl,  amsmath,amssymb,  reprint,]{revtex4-1}%
\usepackage{graphicx}
\usepackage{dcolumn}
\usepackage{bm}
\usepackage{amsmath}
\usepackage{amsfonts}
\usepackage{amssymb}%
\providecommand{\U}[1]{\protect\rule{.1in}{.1in}}
\graphicspath{
	{"C:/Users/Jiri/Dropbox/Papers/Chemistry_papers/2019/TGA_decoherence/figures/"}
	{"d:/Group Vanicek/Desktop/Nikolay/TGA_decoherence/"}
	{figures/}
}
\begin{document}
\preprint{ }
\title{On-the-fly \textit{ab initio} semiclassical evaluation of electronic
coherences in polyatomic molecules reveals a simple mechanism of decoherence}
\author{Nikolay V. Golubev}
\email{nik.v.golubev@gmail.com}
\author{Tomislav Begu\v{s}i\'{c}}
\author{Ji\v{r}\'{\i} Van\'{\i}\v{c}ek}
\email{jiri.vanicek@epfl.ch}
\affiliation{Laboratory of Theoretical Physical Chemistry, Institut des Sciences et
Ing\'enierie Chimiques, Ecole Polytechnique F\'ed\'erale de Lausanne (EPFL),
CH-1015 Lausanne, Switzerland}
\date{\today}

\begin{abstract}
Irradiation of a molecular system by an intense laser field can trigger
dynamics of both electronic and nuclear subsystems. The lighter electrons
usually move on much faster, attosecond time scale but the slow nuclear
rearrangement damps ultrafast electronic oscillations, leading to the
decoherence of the electronic dynamics within a few femtoseconds. We show that
a simple, single-trajectory semiclassical scheme can evaluate the electronic
coherence time in polyatomic molecules accurately by demonstrating an
excellent agreement with full-dimensional quantum calculations. In contrast to
numerical quantum methods, the semiclassical one reveals the physical
mechanism of decoherence beyond the general blame on nuclear motion. In the
propiolic acid, the rate of decoherence and the large deviation from the
static frequency of electronic oscillations are quantitatively described with
just two semiclassical parameters---the phase space distance and signed area
between the trajectories moving on two electronic surfaces. Because it
evaluates the electronic structure on the fly, the semiclassical technique
avoids the \textquotedblleft curse of dimensionality\textquotedblright\ and
should be useful for preselecting molecules for experimental studies.

\end{abstract}

\pacs{Valid PACS appear here}
\maketitle

Recent progress in laser
technologies~\cite{Krausz_Ivanov:2009,Young_Leone:2018,Mauritsson_Vozzi:2018}
has revolutionized the field of atomic and molecular physics. In particular,
tremendous developments of coherent light sources enabled the creation of
sub-femtosecond laser pulses with remarkably well controlled
parameters~\cite{Krausz:2016}. Using state-of-the-art lasers, one is able to
initiate and probe processes that are driven solely by the electron
correlation, i.e., to study and manipulate electron dynamics on its natural
time scale~\cite{Kraus_Woerner:2015}.

Experimental measurements of the electron motion in isolated atoms were
reported~\cite{Goulielmakis_Krausz:2010,Xie_Kitzler:2012}, whereas a direct
evidence of ultrafast electron dynamics in molecules remains a point of
debate~\cite{Lepine_Vrakking:2014}. In particular, there are contradictions
between recent experimental
studies~\cite{Calegari_Nisoli:2014,Lara-Astiaso_Martin:2018} claiming to have
observed the ultrafast electronic processes in molecules and theoretical
investigations~\cite{Arnold_Santra:2017,Vacher_Malhado:2017} performed on
systems of similar complexity. The disagreement is centered around the
question on how strong is the influence of the slow nuclear motion on the
dynamics of electronic density. Extensive \textit{ab initio} calculations for
several molecules~\cite{Arnold_Santra:2017,Vacher_Malhado:2017} demonstrated
that the nuclear dynamics leads to the decoherence of the electronic wave
packet on the time scale of a few femtoseconds which can make experimental
observations of the electronic motion impossible. At the same time,
long-lasting electronic coherences were reported for the ionized propiolic
acid~\cite{Despre_Kuleff:2018} and iodoacetylene~\cite{Jia_Yang:2019},
suggesting that the influence of nuclear motion on the electronic dynamics is
very case-specific and requires careful investigation.

Understanding the interplay between the nuclear rearrangement and ultrafast
electronic motion requires a concerted description of the electron-nuclear
dynamics. Being one of the most powerful approaches for this purpose, the
multi-configurational time-dependent Hartree (MCTDH)
method~\cite{Meyer_Cederbaum:1990,Raab_Cederbaum:1999,Beck_Meyer:2000} was
recently applied for describing electronic
coherence~\cite{Arnold_Santra:2017,Despre_Kuleff:2018}. Although this rigorous
technique takes into account all quantum effects, such as tunneling and
non-adiabatic transitions, it suffers from exponential scaling and also
requires constructing global potential energy surfaces (PESs).

An alternative strategy for simulating coupled electron-nuclear dynamics
employs a trajectory-guided Gaussian basis to represent the evolving
wavepacket and an \textquotedblleft on-the-fly\textquotedblright\ evaluation
of the electronic structure. These \textquotedblleft direct
dynamics\textquotedblright\ approaches calculate the PESs along trajectories
only, thus avoiding the precomputation of globally fitted surfaces, and sample
only the relevant regions of the configuration space. Among these methods, the
closest in spirit to MCTDH are the variational multi-configurational Gaussians
(vMCG)~\cite{Burghardt_Meyer:1999,Worth_Burghardt:2003,Worth_Lasorne:2008,Sulc_Vanicek:2013}%
, but many others exist, ranging from \textit{ab initio} multiple
spawning~\cite{Curchod_Martinez:2018} and other Gaussian basis
methods~\cite{Shalashilin_Child:2004,Saita_Shalashilin:2012} to more
approximate mixed quantum-classical \cite{Tully:1990,Li_Frisch:2005} and
semiclassical \cite{Herman_Kluk:1984,Tatchen_Pollak:2009,Ceotto_Atahan:2009a} approaches.

Treating electronic coherence with trajectory-based techniques was pioneered
by Bearpark, Robb and
co-workers~\cite{Vacher_Robb:2015,Vacher_Robb:2015a,Jenkins_Robb:2016,Jenkins_Robb:2016a}
who propagated multiple trajectories representing an initially delocalized
wave packet using Ehrenfest approximation. Although this technique captures
decoherence due to a superposition of coherent oscillations with different
frequencies appearing at the respective nuclear geometries, it completely
ignores the decoherence due to the quantum motion of the wave packet resulting
in the accumulation of phase along the propagated trajectory. The latter
mechanism is referred to as the phase jitter~\cite{Fiete_Heller:2003} in
general literature on quantum decoherence or, more specifically, as dephasing
in the case of electron-nuclear processes. Allowing the wave packet to evolve
quantum mechanically, dephasing mechanism was taken into account and the
electronic coherence upon ionization of a system was simulated in several
molecules using the direct dynamics version of vMCG
scheme~\cite{Vacher_Malhado:2017} and its Ehrenfest-based
variant~\cite{Jenkins_Robb:2018}.

Here, we use the thawed Gaussian approximation
(TGA)~\cite{Heller:1975,Begusic_Vanicek:2018a,Lasser_Lubich:2020}, one of the
simplest semiclassical approaches for molecular dynamics, to evaluate the
influence of nuclear motion on the ultrafast electronic dynamics and to find a
simple, yet detailed mechanism of decoherence, which is not available in basis
set methods such as MCTDH\ or vMCG. Within the TGA, the nuclear wavefunction
is described by a single Gaussian wave packet whose center follows Hamilton's
equations of motion and whose time-dependent width and phase are propagated
using the local harmonic approximation of the PES.

We simulate coupled electron-nuclear dynamics taking place after outer-valence
ionization of two polyatomic molecules: propiolic acid (HC$_{2}$COOH) and its
amide derivative propiolamide (HC$_{2}$CONH$_{2}$). Propiolic acid provides us
a perfect system for validating the semiclassical TGA because the electronic
coherences in this molecule were recently calculated using a full quantum
MCTDH approach~\cite{Despre_Kuleff:2018}. The propiolamide molecule, in turn,
is studied here for the first time.

The starting point of our investigations is a neutral molecule in its ground
electronic and vibrational states. The ionization of the system performed by
the ultrashort laser pulse can bring the molecule to a non-stationary
superposition of ionic states, thus launching a coupled dynamics of electronic
and nuclear wave packets. A coherent superposition of multiple electronic
states triggers oscillations of the charge along a molecular chain. This
purely electronic mechanism was termed \textit{charge migration}%
~\cite{Cederbaum_Zobeley:1999,Kuleff_Cederbaum:2014} to distinguish it from a
more common charge transfer driven by nuclei~\cite{Sun_Remacle:2017}. Although
the charge migration is governed by the electronic motion, it is strongly
coupled with the nuclear dynamics and therefore can crucially affect the
behavior of the whole molecule.

Previous calculations of the ionization spectrum of propiolic acid
showed~\cite{Golubev_Kuleff:2017,Despre_Kuleff:2018} that, due to the electron
correlation, the ground and the second excited ionic states of the molecule
are a strong mixture of two one-hole configurations: an electron missing in
the highest occupied molecular orbital (HOMO) and an electron missing in the
HOMO--2. Therefore, a sudden removal of an electron either from HOMO or from
HOMO--2 will create an electronic wave packet, which will initiate charge
migration oscillations between the carbon triple bond and the carbonyl oxygen
with a period of about 6.2~fs, determined by the energy gap between the first
and the third cationic states~\cite{Golubev_Kuleff:2017}. Due to the planar
geometry of the propiolic acid and a large energy gap between remaining ionic
states, the indicated superposition can be obtained in an experiment by an
appropriate orientation of the molecule with respect to the laser polarization.

\begin{figure}
\includegraphics[width=8.5cm]{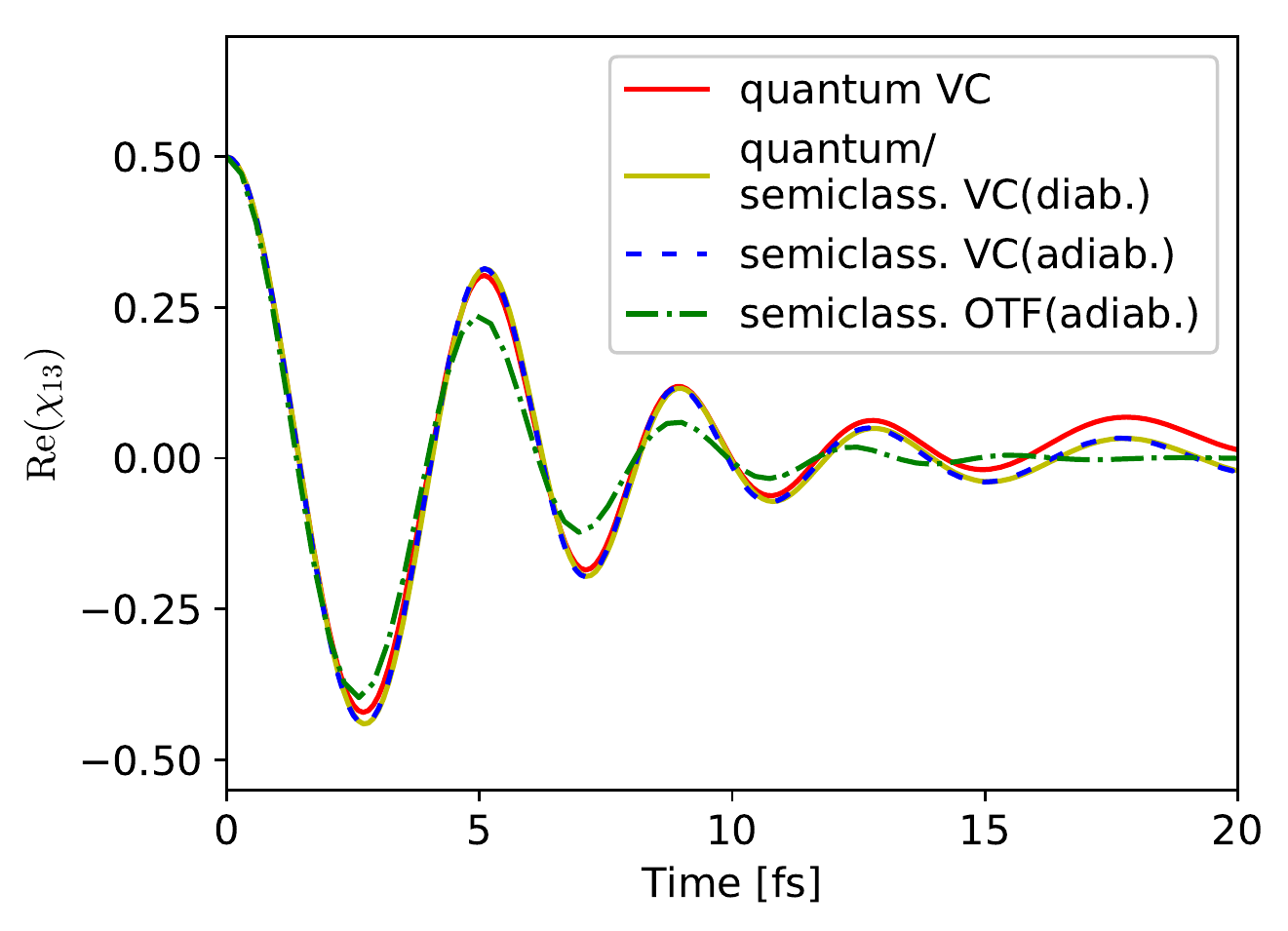}
\caption{Electronic coherence measured by the time-dependent overlap $\chi_{13}(t)$ of the nuclear wave
packets propagated on the first and third cationic states of propiolic acid after removal of a HOMO electron.
Dynamics was performed with the full quantum MCTDH method (``quantum'') or with the semiclassical
vertical-Hessian thawed Gaussian approximation (``semiclass.''), and either with the vibronic coupling (``VC'')
or on-the-fly (``OTF'') Hamiltonian, both obtained at the \textit{ab initio} many-body Green's function ADC(3) level
of theory. All simulations except the ``quantum VC'' calculation employed the diabatic approximation (``diab.''), which
neglects nonadiabatic couplings between the diabatic states, or the adiabatic approximation (``adiab.''), which
neglects nonadiabatic couplings between adiabatic states.}
\label{fig:prop_acid_coherences}
\end{figure}

To describe the ionization process, we used the non-Dyson
algebraic-diagrammatic-construction (ADC) scheme~\cite{Schirmer_Stelter:1998}
to represent the one-particle Green's function. We chose a rather
computationally expensive ADC method because it allows treating the ionization
process explicitly starting from the neutral state of a system. This important
advantage over other, more conventional electronic structure approaches allows
us to estimate populations of the resulting ionic states created after removal
of an electron from a molecule. [See Sec.~I Supplementary Material (SM) for details.]

Within the sudden and Franck--Condon approximations, the ionization is
modelled by projecting the ground (electronic and nuclear) neutral state of a
molecule onto the ionic subspace of the system. After ionization, a single
nuclear Gaussian wave packet on each involved ionic surface is propagated
independently from the others (non-adiabatic effects are neglected). The
center of each Gaussian is guided by a single classical trajectory, while the
width and phase are propagated using the
single-Hessian~\cite{Begusic_Vanicek:2019} variant of the TGA.

Within the Born--Huang representation~\cite{book_Born_Huang:1954} of the
molecular wavefunction, the expectation value of an electronic operator
$\hat{O}(\mathbf{r},\mathbf{R})$ can be expressed as
\begin{equation}
\langle\hat{O}\rangle(t)=\sum_{i,j}\int\chi_{i}^{\ast}(\mathbf{R}%
,t)O_{ij}(\mathbf{R})\chi_{j}(\mathbf{R},t)d\mathbf{R},
\label{eq:expect_value}%
\end{equation}
where $O_{ij}(\mathbf{R})=\int\Phi_{i}^{\ast}(\mathbf{r},\mathbf{R})\hat
{O}(\mathbf{r},\mathbf{R})\Phi_{j}(\mathbf{r},\mathbf{R})d\mathbf{r}$ denote
the $\mathbf{R}$-dependent matrix elements of the electronic operator between
electronic states $i$ and $j$, and the quantities $\chi_{i}(\mathbf{R},t)$ are
the time-dependent nuclear wave packets propagated on the corresponding PESs.

If both the operator $\hat{O}(\mathbf{r},\mathbf{R})$ and the electronic
states $\{\Phi_{i}(\mathbf{r},\mathbf{R})\}$ depend weakly on nuclear
coordinates $\mathbf{R}$, Eq.~(\ref{eq:expect_value}) can be further
simplified as
\begin{equation}
\langle\hat{O}\rangle(t)\approx\sum_{i,j}O_{ij}\chi_{ij}(t),
\label{eq:expect_value_expanded}%
\end{equation}
where the nuclear overlaps $\chi_{ij}(t)=\int\chi_{i}^{\ast}(\mathbf{R}%
,t)\chi_{j}(\mathbf{R},t)d\mathbf{R}$ represent the populations of electronic
states when $i=j$ and the electronic
coherences~\cite{Arnold_Santra:2017,Vacher_Malhado:2017,Despre_Kuleff:2018}
when $i\neq j$. Equation~(\ref{eq:expect_value_expanded}) shows that the
factors $\chi_{ij}(t)$ provide the only source of time dependence in the
expectation value of the electronic operator and thus can serve as convenient
properties to quantify the decoherence time. Electronic coherence evaluated as
the nuclear overlap $\chi_{ij}(t)$ is a special case of \emph{fidelity
amplitude} (also called Loschmidt echo), a quantity measuring sensitivity of
quantum dynamics to perturbations and defined as the overlap at time $t$ of
two wavepackets, initially the same but propagated with different Hamiltonians
\cite{Peres:1984,Jalabert_Pastawski:2001,Vanicek:2006,Gorin_Znidaric:2006,Jacquod_Petitjean:2009,Mollica_Vanicek:2011,Zambrano_Almeida:2011,Goussev_Wisniacki:2012}%
.

We computed the electronic coherences $\chi_{ij}(t)$ generated after ionizing
an electron from the HOMO of the propiolic acid; the initial molecular wave
packet was an equally weighted and phase-synchronized superposition of the
first and third cationic states. Figure~\ref{fig:prop_acid_coherences} shows
the electronic coherence evaluated by five different schemes. We adopted the
vibronic-coupling (VC) Hamiltonian from Ref.~\cite{Despre_Kuleff:2018} to
perform MCTDH simulations taking into account all nuclear degrees of freedom.
The full quantum-mechanical calculations show (red solid line in
Fig.~\ref{fig:prop_acid_coherences}) that the electronic oscillations are
strongly influenced by nuclear motion---the coherences are completely
suppressed within first 15~fs~\cite{Despre_Kuleff:2018}.

To validate the applicability of the TGA, we performed semiclassical
calculations using \textit{adiabatic} version of the VC Hamiltonian, where the
PESs were obtained by diagonalizing the four-state VC model used in the MCTDH
calculations. Our simulation shows (blue dashed line in
Fig.~\ref{fig:prop_acid_coherences}) that on the short time scale the TGA
gives results almost identical to the full quantum MCTDH calculations. The
small deviations start to appear at longer times due to the nonadiabatic
effects, neglected in the TGA. Note that TGA is exact within the VC model when
the diabatic PESs are not coupled to each other, and therefore provides
results identical to the MCTDH simulations performed on such VC Hamiltonian
(yellow solid line in Fig.~\ref{fig:prop_acid_coherences}).

\begin{figure}
\includegraphics[width=\columnwidth]{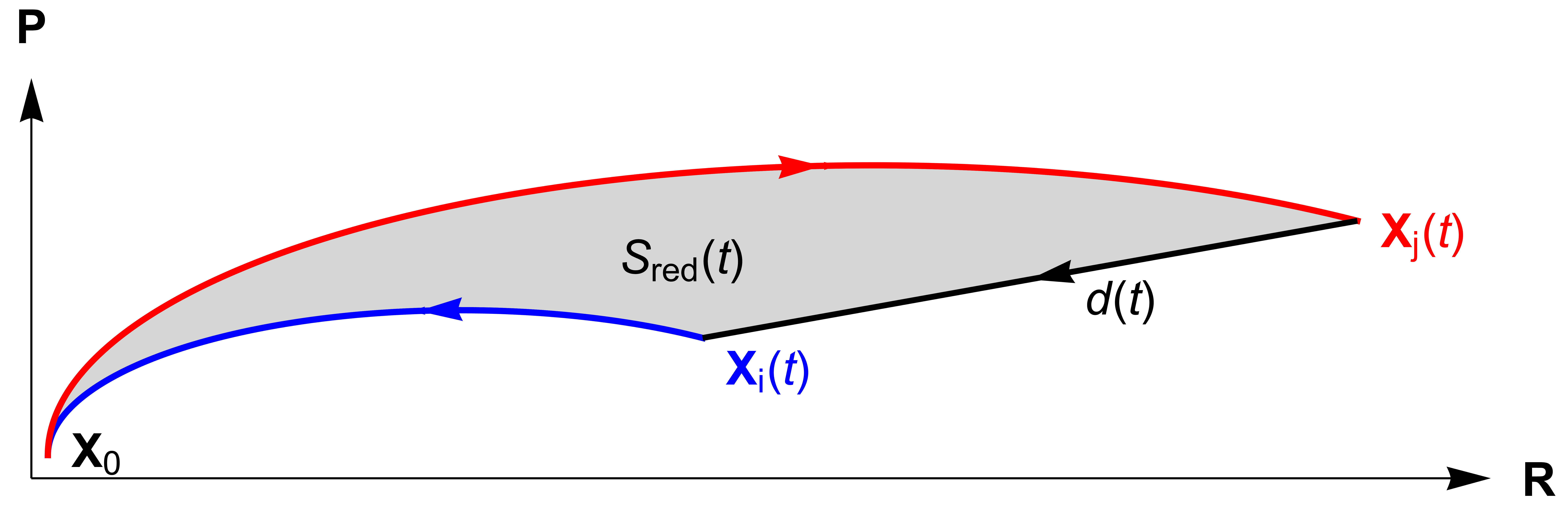}
\caption{Geometric interpretation of Eq.~(\ref{eq:S_red}): The reduced action $S_{\text{red}}$ is the gray
phase-space area enclosed by the curve $\mathbf{C}$ consisting of the classical trajectory propagated
in state $i$ with potential energy $V_i$ and connecting
$\mathbf{X}_i(t)$ with $\mathbf{X}_0$ (blue curve), classical trajectory propagated in state $j$ with
potential energy $V_j$ and connecting $\mathbf{X}_0$ with
$\mathbf{X}_j(t)$ (red curve), and a straight (black) line connecting $\mathbf{X}_j(t)$ with $\mathbf{X}_i(t)$.
The correct sign of $S_{\text{red}}$ is obtained by taking the curve integral along $\mathbf{C}$ in the direction
indicated by the arrows. The phase space distance betwen the final points $\mathbf{X}_j(t)$ and $\mathbf{X}_i(t)$,
given by the  length $d$ of the black line segment, determines the coherence decay [i.e., the decay of
$|\chi_{ij}(t)|$], while the reduced action $S_{\text{red}}$, equal to the gray area, affects the frequency of
oscillations of the coherence $\chi_{ij}(t)$.} \label{fig:S_red}
\end{figure}

We also performed semiclassical calculations with on-the-fly evaluation of the
electronic structure at the same \textit{ab initio} level of theory as that
used in the construction of the VC Hamiltonian (green dash-dotted line in
Fig.~\ref{fig:prop_acid_coherences}). In this case, the wave packet can
potentially evolve beyond a simple model used for fitting PESs. In particular,
VC Hamiltonian typically uses a rather primitive approximation of PESs for
nuclear configurations formed by superposition of normal modes. Allowing the
wave packet to evolve according to the exact Hamiltonian computed on the fly
makes it possible to visit nuclear regions inaccessible within the VC
Hamiltonian and thus to take anharmonicity effects into account (see Sec.~II
of SM for details). This is reflected in our on-the-fly calculations, which
predict the electronic motion with a similar oscillation period, but a
slightly faster decay of the electronic coherence than within the VC model.
Remarkably, because the effect of using the on-the-fly potential is much
larger than the effect of including the nonadiabatic couplings, the
semiclassical on-the-fly result of the TGA\ is most likely more accurate than
the quantum result of the MCTDH calculation with the VC model!

In Sec.~III of SM, we derive an analytical expression%
\begin{equation}
\chi_{ij}(t)=e^{-d(t)^{2}/4\hbar}e^{iS(t)/\hbar} \label{eq:chi_12_res}%
\end{equation}
for the semiclassically evaluated coherence in case the two Gaussians have
fixed widths. Here $d(t)$ is phase-space distance (in mass- and
frequency-scaled phase-space coordinates $\mathbf{R}$ and $\mathbf{P}$)
between the centers of the two Gaussian wavepackets at time $t$,
\begin{equation}
S(t)=S_{\text{red}}(t)-\Delta E\,t, \label{eq:S_t_res}%
\end{equation}
is the classical action, $\Delta E=V_{j}(\mathbf{R}_{0})-V_{i}(\mathbf{R}%
_{0})$ is the energy gap at the initial point $\mathbf{R}_{0}$, and
\begin{equation}
S_{\text{red}}(t)=\oint_{\mathbf{C}}\mathbf{P}^{T}\cdot d\mathbf{R}
\label{eq:S_red}%
\end{equation}
is the reduced action equal to the signed area within the closed curve
$\mathbf{C}$ shown in Fig.~\ref{fig:S_red}\ \cite{Zambrano_Almeida:2011}.

\begin{figure}
\includegraphics[width=8.5cm]{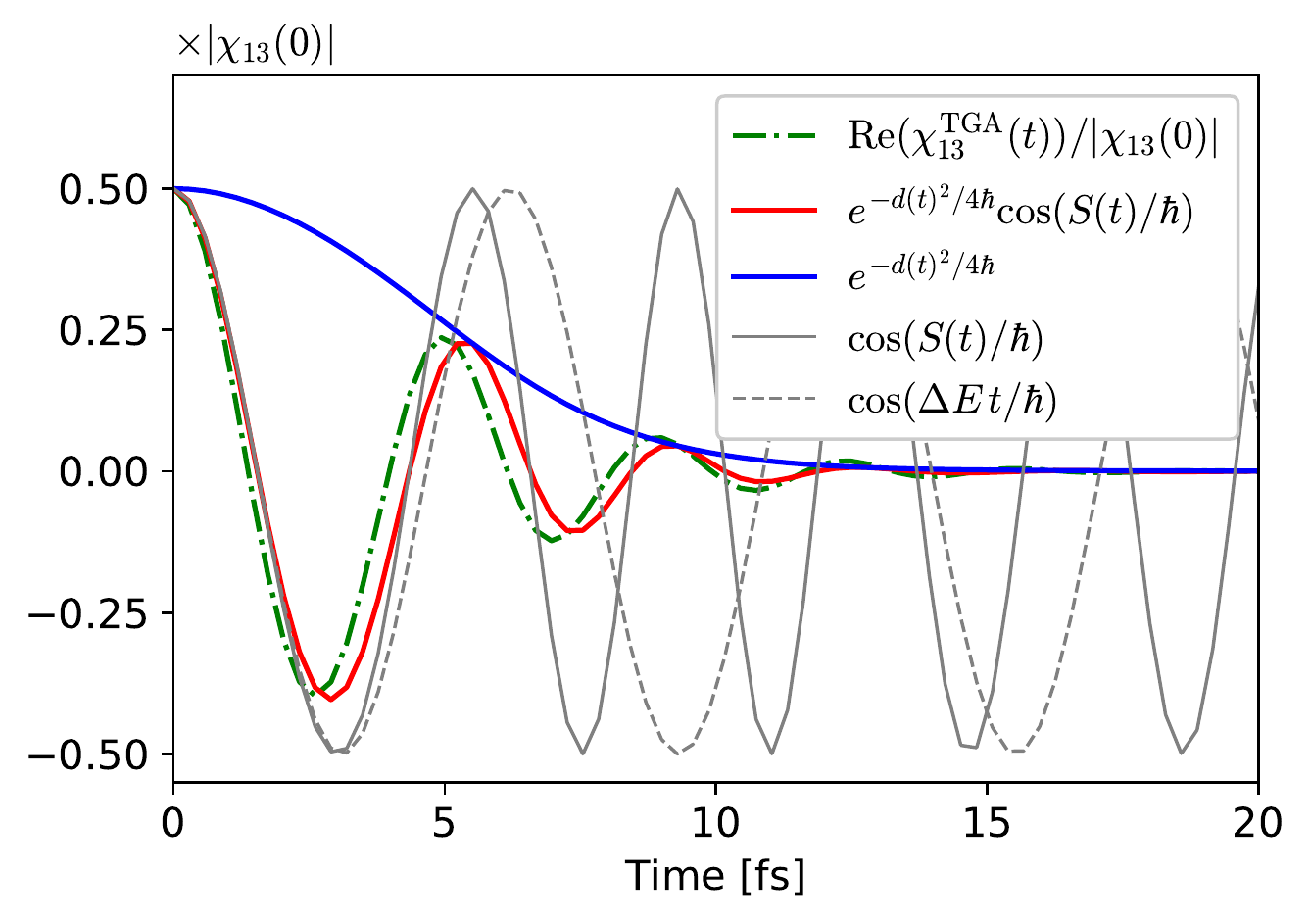}
\caption{Semiclassical analysis of the electronic coherence $\chi_{13}(t)$ from
Fig.~\ref{fig:prop_acid_coherences}. Comparison of the coherence computed with the on-the-fly version of TGA
(green dash-dotted line), analytical semiclassical
expression~(\ref{eq:chi_12_res}) (red solid line), factors describing the decay (blue solid line)
and oscillations (gray solid line) of coherence in the presence of nuclear motion, and the undamped coherence
in the absence of nuclear motion (gray dashed line).}
\label{fig:coherence_interpretation}
\end{figure}

The analytical expression (\ref{eq:chi_12_res}) provides a simple,
semiclassical interpretation of the effect of nuclear dynamics on electronic
coherence (see Fig.~\ref{fig:coherence_interpretation}): The diverging nuclear
trajectories affect not only the absolute value of $\chi_{ij}(t)$, which, as
expected, decays as a Gaussian function of the phase-space distance $d(t)$,
but also frequency of electronic oscillations. In the absence of nuclear
motion, the electronic coherence would oscillate with frequency $\Delta
E/\hbar$, but now, due to nuclear dynamics, the phase of electronic
oscillations at time $t$ is modified by the area $S_{\text{red}}(t)$ divided
by $\hbar$ [see Eqs.~(\ref{eq:chi_12_res}), (\ref{eq:S_t_res}), and
(\ref{eq:S_red})]. It is easy to see that if the potential energy surfaces are
simply vertically shifted, i.e., if $V_{j}=V_{i}+\Delta E$, then
$\mathbf{X}_{i}=\mathbf{X}_{j}$ and $d=S_{\text{red}}=0$, implying that the
electronic coherence $\chi_{ij}(t)$ is not affected by nuclear dynamics.

Let us turn to the electron-nuclear dynamics driven by the ionization of the
propiolamide. Similar to the spectrum of the propiolic
acid~\cite{Golubev_Kuleff:2017}, in the energy range 10--14 eV only the four
states shown in Fig.~2\ in Sec.~IV of the SM are present. The strong electron
correlation between valence orbitals in the neutral propiolamide leads to
appearance of the almost equal in weights one-hole configurations in the ionic
states. Therefore, an ultrashort (sudden) ionization of the molecule will
inevitably create an electronic wave packet and trigger dynamics of the
electron density between the carbon triple bond and amide moiety. The molecule
is planar and thus belongs to the $C_{s}$ symmetry group which allows
assignment of the ionic states to two irreducible representations: the second
and fourth states belong to the A$^{\prime}$, while the first and third states
to A$^{\prime\prime}$. As for propiolic acid, orientation of the propiolamide
with respect to polarization of the ionizing laser field can be used to
populate only the states of interest.

\begin{figure}
\includegraphics[width=8.5cm]{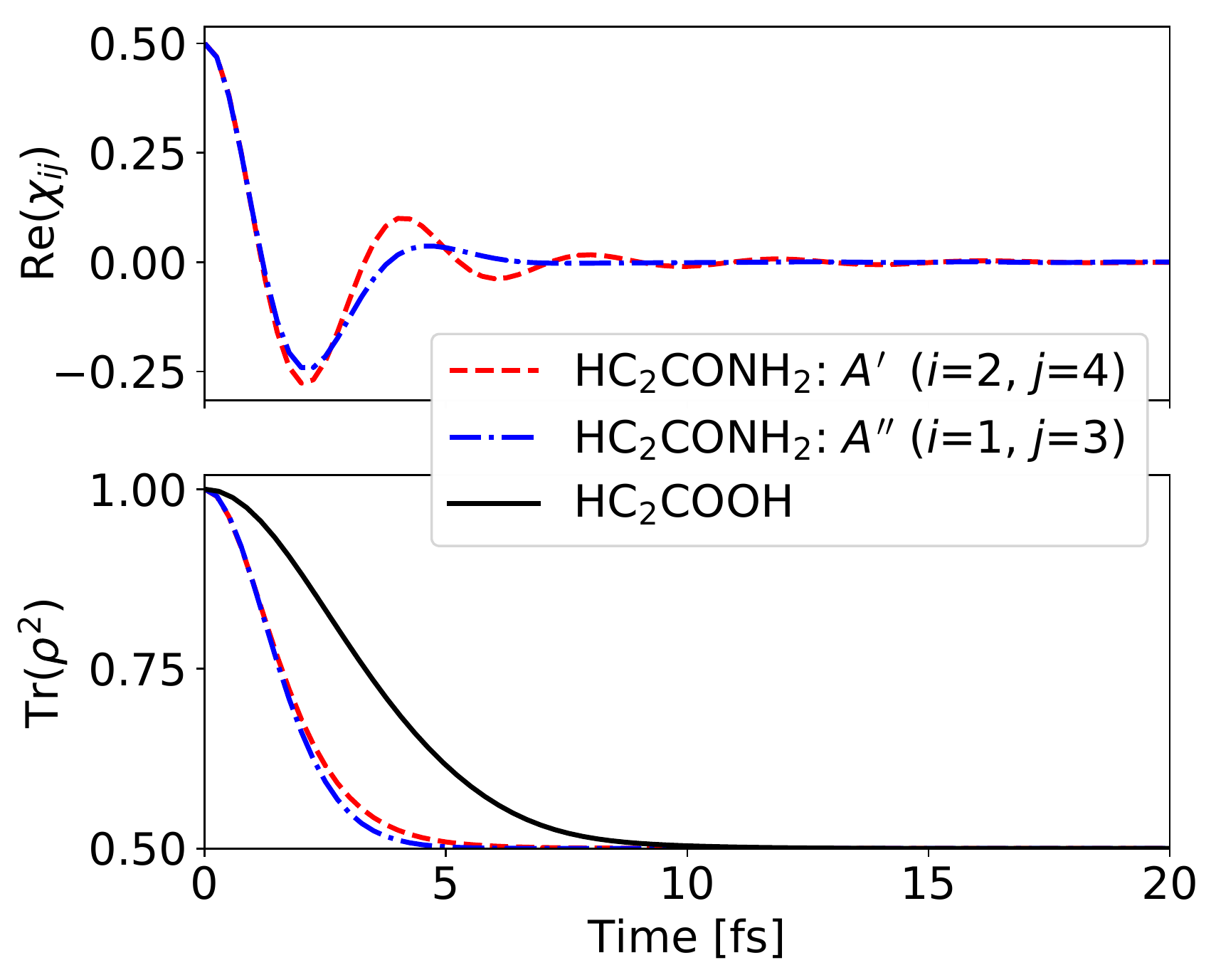}
\caption{Top panel: Time evolution of the electronic coherences $\chi_{ij}(t)$ created by the equally weighted
coherent superposition of the second and fourth cationic states (ionization from the HOMO--1, A$'$ symmetry),
and the first and third cationic states (ionization from the HOMO, A$''$ symmetry). Bottom panel: Comparison of
the electronic purity function $\text{Tr}(\rho^2)$ for the propiolamide and propiolic acid molecules.}
\label{fig:propiolamide_coherences}
\end{figure}

Figure~\ref{fig:propiolamide_coherences} shows the evolution of the electronic
coherence in the propiolamide. Taking advantage of the molecular symmetry, we
simulate dynamics occurring after removal of the HOMO (populating the first
and third cationic states) and HOMO--1 (populating the second and fourth
cationic states) electrons. Starting from the equally weighted and
phase-matched superposition of the electronic states, in both cases the
nuclear motion perturbs electronic oscillations and leads to decoherence.
Coherence times of different molecules can be compared using the purity
$\operatorname{Tr}(\rho(t)^{2})$~\cite{Arnold_Santra:2017}, where the
electronic density matrix $\rho(t)$ is related to the matrix of nuclear
overlaps from Eq.~(\ref{eq:expect_value_expanded}) by transposition:
$\rho_{ij}(t)=\chi_{ji}(t)$. Due to decoherence, the purity decays from the
value $\text{Tr}(\rho(0)^{2})=1$ for the initially pure state to the value
$1/n$ for the equally weighted mixture of $n$ states.

Our simulations demonstrate that, contrary to the propiolic acid molecule, for
which long-lasting coherences were observed (see bottom panel of
Fig.~\ref{fig:propiolamide_coherences} and also Ref.~\cite{Despre_Kuleff:2018}%
), the initially pure superpositions in the propiolamide evolve to mixed
states in just a few femtoseconds. Importantly, the energy gaps between the
involved electronic states of propiolamide are larger than those for the
propiolic acid (see Fig.~2 of SM and Ref.~\cite{Golubev_Kuleff:2017}), which
leads to faster oscillations of electronic density along the molecular chain.
Despite the rather short coherence time, due to the faster charge migration
the electronic density in the propiolamide has enough time to perform one
clear oscillation (see Fig.~\ref{fig:propiolamide_coherences}\ above and
Fig.~3 of SM). Moreover, the existence of the strong hole-mixing in both
symmetries of the propiolamide can be used to induce oscillations of the
charge along different directions in the molecule. Dependence of the charge
migration on the molecular orientation provides an important advantage for
experimental measurements utilizing the time-resolved high-harmonic generation
(HHG) spectroscopy employed recently by W\"{o}rner and
co-workers~\cite{Kraus_Woerner:2015}. Alignment of the molecule with respect
to the pump pulse should be reflected in the resulting HHG spectra and thus
can be used as a direct evidence of the ultrafast electron dynamics.

Although trajectory-based direct dynamics methods were previously used to
estimate electronic coherences in various polyatomic
molecules~\cite{Vacher_Malhado:2017,Jenkins_Robb:2018}, converged results
typically required a large number of trajectories. The vMCG used in these
studies take into account non-adiabatic transitions and tunneling effects, but
require solving rather complicated equations of motion and make interpreting
obtained results less intuitive than the simple picture provided here in
Eq.~(\ref{eq:chi_12_res}) and Figs.~\ref{fig:S_red} and
\ref{fig:coherence_interpretation}. Our approach based on the TGA can be
viewed as a very special case of far more general vMCG---namely, it can be
classified as a single-Gaussian, non-variational, multi-set, non-frozen,
adiabatic, single-Hessian version of vMCG. In contrast to the application of
vMCG in Refs.~\cite{Vacher_Malhado:2017,Jenkins_Robb:2018}, our implementation
of the TGA uses a multi-set approach, where a \textit{single} Gaussian
function with \textit{relaxed} parameters is used for every involved
electronic state. By approximately taking into account quantum properties of
the wave packet, such as its width and phase, the TGA captures the dephasing
mechanism while maintaining sufficient accuracy, especially at short time
scales. A detailed comparison of the TGA and vMCG is provided in Sec.~V of
SM---remarkably, the TGA, which uses only a single classical trajectory per
electronic state, yields in the propiolic acid better results than the
single-set version of vMCG with 31 variational trajectories (see Fig.~4 in
SM). While the multi-set version of vMCG gives similar results to TGA also
with 1 Gaussian, no improvement is seen by using 8 Gaussians (Fig.~4 of SM).
In the TGA, using only two classical trajectories was not only sufficient but
also crucial for revealing the simple physical mechanism of decoherence.

The semiclassical vertical-Hessians TGA used in this paper can be further
improved by calculating Hessians along the propagated trajectory and thus
taking into account more complicated situations, e.g., dissociation of a
molecule. Extensions of the TGA, such as the extended thawed Gaussian
approximation~\cite{Lee_Heller:1982,Patoz_Vanicek:2018} or Hagedorn
wavepackets~\cite{Hagedorn:1998,Lasser_Lubich:2020}, which propagate a
Gaussian multiplied by a linear or general polynomial, can make on-the-fly
semiclassical simulations even more accurate.

In conclusion, we implemented a simple and efficient on-the-fly semiclassical
approach to understand the effects of nuclear motion on electronic coherence
in molecules. Although the propiolic acid and propiolamide have very similar
ionization spectra, our calculations predict that their electronic coherence
times differ substantially. The simple method was validated by comparison with
the full-dimensional quantum calculations performed using the MCTDH and vMCG
methods. As suggested by Fig.~\ref{fig:prop_acid_coherences}, neglecting the
nonadiabatic couplings by the TGA may not be a severe approximation because,
even in systems with strong couplings, the nonadiabatic effects typically
start playing a significant role only at times longer than the ultrashort
decoherence time scale. If one suspects an exceptional importance of
nonadiabatic effects even at times before the electronic coherence has decayed
to zero, it is possible to validate the applicability of the TGA without
performing expensive quantum simulations by verifying adiabaticity with
on-the-fly semiclassical calculations based, e.g., on surface hopping
\cite{Tully:1990} or multiple-surface dephasing representation
\cite{Zimmermann_Vanicek:2010,Zimmermann_Vanicek:2012,Zimmermann_Vanicek:2012a,Prlj_Vanicek:2020}%
.

Despite its limitations, the presented technique for evaluating
coherence\textbf{ }can help breaking the \textquotedblleft curse of
dimensionality\textquotedblright\ appearing in the quantum treatment of large
molecules, which can be crucial for full-dimensional simulations of ultrafast
electronic processes in biologically relevant
systems~\cite{Calegari_Nisoli:2016}. Being able to treat molecules with a few
hundred atoms, this technique can help shed light on the continuing debates on
the role of quantum coherence in biology~\cite{Lambert_Nori:2013}, quickly
preselect molecules suitable for further experimental investigations, and
support theoretically recent experimental observations of attosecond electron
dynamics in realistic molecular systems.

\begin{acknowledgments}
Authors thank Alexander Kuleff for many valuable discussions and the Swiss
National Science Foundation for financial support through the NCCR MUST
(Molecular Ultrafast Science and Technology) Network. N.~V.~G. acknowledges
the support by the Branco Weiss Fellowship---Society in Science, administered
by the ETH Z\"{u}rich.
\end{acknowledgments}


%

\end{document}


\title{Supplementary Material for \textquotedblleft On-the-fly \textit{ab initio}
semiclassical evaluation of electronic coherences in polyatomic molecules
reveals a simple mechanism of decoherence\textquotedblright}
\author{Nikolay V. Golubev}
\email{nik.v.golubev@gmail.com}
\author{Tomislav Begu\v{s}i\'{c}}
\author{Ji\v{r}\'{\i} Van\'{\i}\v{c}ek}
\email{jiri.vanicek@epfl.ch}
\affiliation{Laboratory of Theoretical Physical Chemistry, Institut des Sciences et
Ing\'enierie Chimiques, Ecole Polytechnique F\'ed\'erale de Lausanne (EPFL),
CH-1015 Lausanne, Switzerland}

\begin{abstract}
In the Letter, we have shown how the thawed Gaussian
approximation~(TGA)~\cite{Heller:1975,Begusic_Vanicek:2018a} can be used to
compute electronic coherence time in polyatomic molecules. Here, we present
additional details of the calculations and results mentioned in the main text.
After describing details of electronic structure computations in Sec.~I, in
Sec.~II we discuss advantages of the on-the-fly approach over the previously
reported fully quantum calculations of the electron-nuclear dynamics in the
propiolic acid molecule~\cite{Despre_Kuleff:2018} performed using the
multi-configurational time-dependent
Hartree~(MCTDH)~\cite{Meyer_Cederbaum:1990,Beck_Meyer:2000} method on the
vibronic coupling~(VC) model. In Sec.~III, we derive a simple semiclassical
expression providing an intuitive phase-space interpretation of the
decoherence presented in the main text. Section~IV contains further details of
the calculations of decoherence in the propiolamide. Finally, in Sec.~V we
include a detailed comparison of the TGA with the variational
multi-configurational Gaussians~(vMCG)
method~\cite{Burghardt_Meyer:1999,Worth_Burghardt:2003}, which was recently
reviewed in Ref.~\cite{Richings_Lasorne:2015} and applied to study electronic
coherence in several molecules~\cite{Vacher_Malhado:2017,Jenkins_Robb:2018}.

\end{abstract}
\date{\today}
\maketitle

\section{Details of electronic structure calculations}

Ionization potentials of propiolic acid and propiolamide molecules were
calculated using the third-order ADC [ADC(3)]
scheme~\cite{Schirmer_Stelter:1998} consistent with the exact Green's function
up to the third order of perturbation theory with respect to the Hartree--Fock
reference Hamiltonian. Standard double-zeta plus polarization (DZP) basis
sets~\cite{CanalNeto_Jorge:2005} were employed to construct the noncorrelated
reference states. Ground-state geometries of the neutral molecules were
optimized using the density functional theory~\cite{Hohenberg_Kohn:1964} with
the B3LYP functional~\cite{Stephens_Frisch:1994}. The optimization was done
with Gaussian 16 package~\cite{g16}.

\section{Comparison of trajectories propagated using the vibronic
coupling\ and on-the-fly\ Hamiltonians}

\begin{figure}
\includegraphics[width=0.9\textwidth]{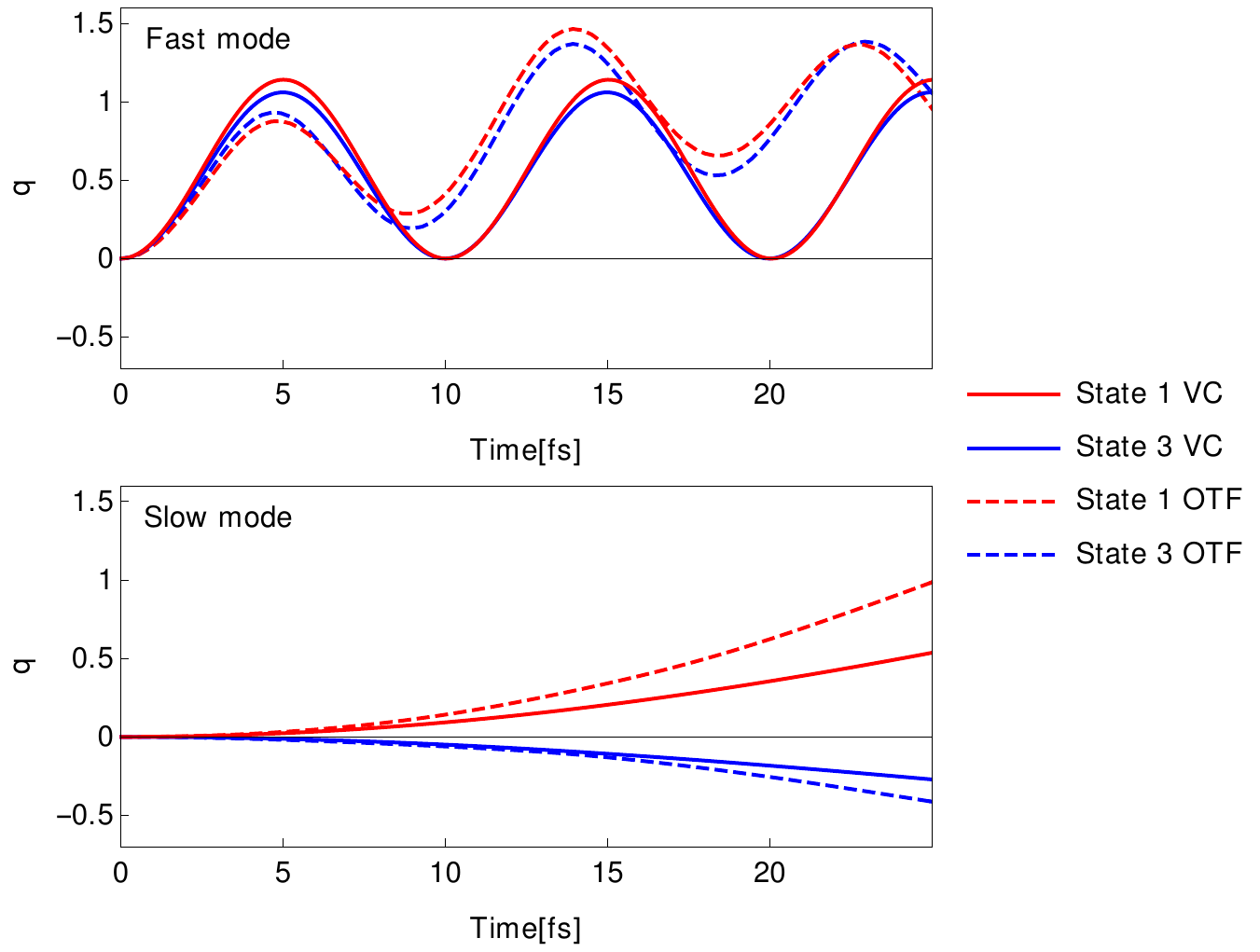}
\caption{Classical trajectories followed by the centers of Gaussian wave packets propagated on the fly (OTF, dashed
lines) or using the vibronic coupling Hamiltonian (VC, solid lines) on the potential energy surface associated with the
first (red color) or third (blue color) cationic state. Normal mode coordinates of these trajectories are shown only for
the fastest (top panel) and slowest (bottom panel) modes of the propiolic acid molecule. The fast mode, despite its
richer dynamics, has a very limited effect on the decay of the coherence because the nuclear wave packets exhibit
almost identical dynamics in both states. In contrast, the wave packets propagated in the two states move to opposite
directions along the slow mode which leads to decoherence.}
\label{fig:normal_modes}
\end{figure}

Whereas the direct dynamics version of the thawed Gaussian approximation (TGA)
takes into account the anharmonicity at least partially, the VC Hamiltonian
employs the harmonic approximation for fitting the potential energy surfaces
(PESs). Clearly, this approximate treatment of the PESs can lead to
substantial differences between \textquotedblleft real\textquotedblright\ and
approximate dynamics. Moreover, vibronic coupling (VC) Hamiltonian utilizes a
rather primitive representation of PESs for nuclear configurations located
\textquotedblleft between\textquotedblright\ normal modes (i.e. such
configurations which are formed by superposition of normal modes). The direct
dynamics approach, in turn, operates with forces computed on the fly and thus
makes it possible to visit nuclear regions that may be accidentally missed in
the VC Hamiltonian. To investigate such effects, in
Fig.~\ref{fig:normal_modes} we show trajectories followed by the centers of
Gaussian wave packets propagated on two surfaces associated with the two
electronic states involved in the dynamics.

The top and bottom panels of the figure show evolution of classical
trajectories along the fastest (high frequency) and slowest (low frequency)
normal modes of the system, respectively. The fast mode has almost no effect
on the overall decoherence because the trajectories propagated on the two
involved surfaces are almost the same, regardless whether the calculation uses
the VC or on-the-fly approach. Due to the anharmonicity of the two PESs, the
on-the-fly classical trajectories gradually drift from the equilibrium point,
but this drift is almost the same for the trajectories on the two surfaces
and, as a result, has almost no effect on the overall decoherence. The
situation is different for the slow mode (see bottom panel of
Fig.~\ref{fig:normal_modes}), where trajectories on the two surfaces move in
opposite directions, resulting in fast decoherence of nuclear wave packets
propagating in the two states. Because the two on-the-fly trajectories diverge
faster than the corresponding VC trajectories, the decay of electronic
coherence is faster in the case of on-the-fly propagation than for the
precomputed VC Hamiltonian. Similar analysis can be repeated for all modes and
explains the slightly faster decay of overall coherence in the on-the-fly
calculation observed in Fig.~1 of the main text. Note that such a simple and
intuitive description of decoherence is available only in the multi-set
approach and, as we show below, is further simplified in the TGA setting, with
a single Gaussian on each surface.

\section{Phase-space description of decoherence}

We now derive an analytical semiclassical expression for the time-dependent
coherence, which yields a simple phase-space interpretation of the effect of
nuclear dynamics on the decay rate and frequency of electronic coherence
oscillations. The derivation and interpretation is simplified by further
assuming the width of the Gaussian wavepackets to be fixed.

Let us consider electronic coherence evaluated as the overlap
\begin{equation}
\chi_{12}(t)=\int d\mathbf{R}\chi_{1}^{\ast}(\mathbf{R},t)\chi_{2}%
(\mathbf{R},t) \label{eq:chi_12}%
\end{equation}
of two frozen nuclear Gaussian wavepackets
\begin{equation}
\chi_{i}(\mathbf{R},t)=\det\left(  \frac{\mathbf{m}\cdot\mathbf{\Omega}}%
{\pi\hbar}\right)  ^{1/4}e^{-\frac{1}{2\hbar}[\mathbf{R}-\mathbf{R}%
_{i}(t)]^{T}\cdot\mathbf{m}^{1/2}\cdot\mathbf{\Omega}\cdot\mathbf{m}%
^{1/2}\cdot\lbrack\mathbf{R}-\mathbf{R}_{i}(t)]+\frac{i}{\hbar}\mathbf{P}%
_{i}(t)^{T}\cdot\lbrack\mathbf{R}-\mathbf{R}_{i}(t)]+\frac{i}{\hbar}\gamma
_{i}(t)},\quad i\in\{1,2\}, \label{eq:fga_wp}%
\end{equation}
evolved with two different Hamiltonians $H_{1}$ and $H_{2}$ from the same
initial Gaussian centered at $\mathbf{R}_{0}$ and $\mathbf{P}_{0}=\mathbf{0}$.
Here $\mathbf{\Omega}$ is the matrix of vibrational frequencies in the
electronic ground state, the fixed width matrix is given by $\mathbf{m}%
^{1/2}\cdot\mathbf{\Omega}\cdot\mathbf{m}^{1/2}$ because the initial state was
a vibrational ground state of the ground electronic state, the pair
$(\mathbf{R}_{i},\mathbf{P}_{i})\equiv\mathbf{X}_{i}$ denotes the phase-space
center of the Gaussian, $\gamma_{i}=S_{i}-[\hbar\operatorname{Tr}%
(\mathbf{\Omega})/2]t$ is the time-dependent phase of the wavepacket, and
$S_{i}$ is the classical action along the classical path connecting
$\mathbf{X}_{0}$ and $\mathbf{X}_{i}$. Before computing the overlap $\chi
_{12}(t)$ explicitly, we introduce mass- and frequency-scaled coordinates and
momenta
\begin{align}
\mathbf{R}^{\prime}  &  :=\mathbf{\Omega}^{1/2}\cdot\mathbf{m}^{1/2}%
\cdot\mathbf{R},\label{eq:scaling_R}\\
\mathbf{P}^{\prime}  &  :=\mathbf{\Omega}^{-1/2}\cdot\mathbf{m}^{-1/2}%
\cdot\mathbf{P}, \label{eq:scaling_P}%
\end{align}
(and immediately drop the prime superscripts for simplicity) to write the
wavepacket as
\begin{equation}
\chi_{i}(\mathbf{R},t)=\exp\left\{  -\frac{1}{2\hbar}[\mathbf{R}%
-\mathbf{R}_{i}(t)]^{2}+\frac{i}{\hbar}\mathbf{P}_{i}(t)^{T}\cdot
\lbrack\mathbf{R}-\mathbf{R}_{i}(t)]+\frac{i}{\hbar}\gamma_{i}(t)\right\}  .
\label{eq:fga_wp_scaled}%
\end{equation}
The overlap is then given analytically as
\begin{equation}
\chi_{12}(t)=e^{-d(t)^{2}/4\hbar}e^{iS(t)/\hbar}, \label{eq:chi_12_res}%
\end{equation}
where
\begin{equation}
d(t)=\left\vert \mathbf{X}_{2}(t)-\mathbf{X}_{1}(t)\right\vert =\sqrt
{|\mathbf{R}_{2}(t)-\mathbf{R}_{1}(t)|^{2}+|\mathbf{P}_{2}(t)-\mathbf{P}%
_{1}(t)|^{2}} \label{eq:d}%
\end{equation}
is the phase-space distance between the centers of the two Gaussian
wavepackets and
\begin{align}
S(t)  &  =\frac{1}{\hbar}\left[  (S_{2}-S_{1})-\mathbf{\overline{P}}^{T}%
\cdot\Delta\mathbf{R}\right]  ,\label{eq:S_t}\\
\mathbf{\overline{P}}  &  =\frac{1}{2}(\mathbf{P}_{1}(t)+\mathbf{P}%
_{2}(t)),\label{eq:P_bar}\\
\Delta\mathbf{R}  &  =\mathbf{R}_{2}(t)-\mathbf{R}_{1}(t). \label{eq:delta_R}%
\end{align}
We now rewrite the action $S_{i}$ on each surface as
\begin{align}
S_{i}(t)  &  =\int_{0}^{t}L_{i}d\tau\label{eq:S_i_1}\\
&  =\int_{0}^{t}(\mathbf{P}_{i}^{T}\cdot\dot{\mathbf{R}}-H_{i})d\tau
\label{eq:S_i_2}\\
&  =\int_{\mathbf{R}_{0}}^{\mathbf{R}_{i}(t)}\mathbf{P}_{i}^{T}\cdot
d\mathbf{R}-tV_{i}(\mathbf{R}_{0}). \label{eq:S_i_3}%
\end{align}
$L_{i}$ is the Lagrangian evaluated along the classical path $\mathbf{X}_{i}$,
$H_{i}(\mathbf{X}_{i})=H_{i}(\mathbf{X}_{0})$ because the Hamiltonian is a
constant of motion, and $H_{i}(\mathbf{X}_{0})=V_{i}(\mathbf{R}_{0})$ because
$\mathbf{P}_{0}=\mathbf{0}$. Then,
\begin{equation}
S(t)=S_{\text{red}}(t)-\Delta Et, \label{eq:S_t_res}%
\end{equation}
where
\begin{equation}
\Delta E=V_{2}(\mathbf{R}_{0})-V_{1}(\mathbf{R}_{0}) \label{eq:delta_E}%
\end{equation}
is the energy gap at $\mathbf{R}_{0}$ and
\begin{align}
S_{\text{red}}(t)  &  =\int_{\mathbf{R}_{0}}^{\mathbf{R}_{2}(t)}\mathbf{P}%
_{2}^{T}\cdot d\mathbf{R}-\int_{\mathbf{R}_{0}}^{\mathbf{R}_{1}(t)}%
\mathbf{P}_{1}^{T}\cdot d\mathbf{R}-\mathbf{\overline{P}}^{T}\cdot
\Delta\mathbf{R}\label{eq:S_red_1}\\
&  =\oint_{\mathbf{C}}\mathbf{P}^{T}\cdot d\mathbf{R} \label{eq:S_red}%
\end{align}
is the reduced action equal to the signed area within the closed curve
$\mathbf{C}$ consisting of the two classical paths connecting $\mathbf{X}_{0}$
with $\mathbf{X}_{i}(t)$ ($i\in\{1,2\}$) and a straight line connecting
$\mathbf{X}_{1}$ and $\mathbf{X}_{2}$ (see Fig.~2 of the main text, where the
indices $1$, $2$ are replaced with general indices $i$, $j$)
\cite{Zambrano_Almeida:2011}.

\section{Charge migration in propiolamide molecule}

\begin{figure}
\includegraphics[width=0.9\textwidth]{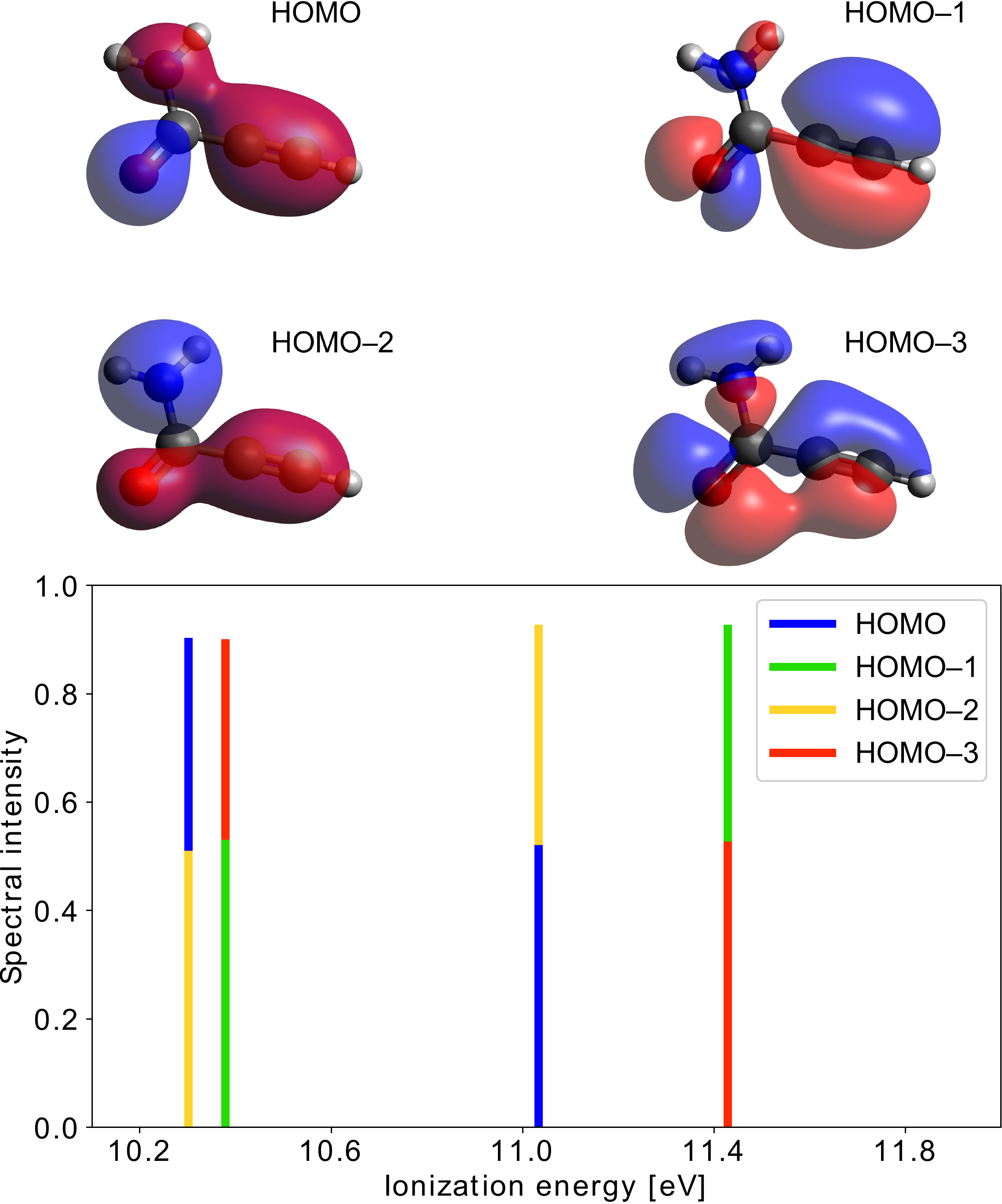}
\caption{First four cationic states of the propiolamide computed using the \textit{ab initio} many-body Green's function
ADC(3) method. The second and fourth states belong to the $A'$ symmetry, while the first and third states to $A''$.
The next ionic state is located at 14.5 eV. The spectral intensity is defined as the combined weight of all one-hole
configurations in the configuration-interaction expansion of the ionic state. The orbitals involved in the hole-mixing
are also depicted.} \label{fig:propiolamide_spec}
\end{figure}

Figure~\ref{fig:propiolamide_spec} shows the first four ionic states of
propiolamide. The next excited cationic state is located at 14.5 eV. The
strong electron correlation between valence orbitals in the neutral
propiolamide leads to appearance of the almost equal in weights one-hole
configurations in the ionic states. Therefore, as for the propiolic acid, an
ultrashort (sudden) ionization of the molecule will inevitably create an
electronic wave packet and trigger dynamics of the electron density between
the carbon triple bond and amide moiety of the system.

One way to describe and visualize the dynamics of electrons along the
molecular chain is to compute the hole density of the system. The latter is
defined as the difference between the electronic density of the system before
ionization (the neutral) and the electronic density of the
cation~\cite{Cederbaum_Zobeley:1999,Breidbach_Cederbaum:2003,Kuleff_Cederbaum:2005}
\begin{equation}
Q(\mathbf{r},t)=\int_{\mathbf{R}}d\mathbf{R}\left(  \langle\Psi^{N}%
(\mathbf{R})|\hat{\rho}(\mathbf{r},\mathbf{R})|\Psi^{N}(\mathbf{R}%
)\rangle-\langle\Psi^{N-1}(\mathbf{R},t)|\hat{\rho}(\mathbf{r},\mathbf{R}%
)|\Psi^{N-1}(\mathbf{R},t)\rangle\right)  , \label{eq:Q_full}%
\end{equation}
where $\hat{\rho}(\mathbf{r},\mathbf{R})$ is the one-body electronic density
operator, $\Psi^{N}(\mathbf{r},\mathbf{R})$ is the stationary ground molecular
state of the neutral, and $\Psi^{N-1}(\mathbf{R},t)$ is the time-dependent
molecular wavefunction of the ion. Assuming that both electronic states and
the density operator $\hat{\rho}(\mathbf{r},\mathbf{R})$ have a weak
dependence on nuclear coordinates $\mathbf{R}$, Eq.~(\ref{eq:Q_full}) can be
further simplified as
\begin{equation}
Q(\mathbf{r},t)\approx\langle\Phi_{0}^{N}|\hat{\rho}(\mathbf{r})|\Phi_{0}%
^{N}\rangle-\sum_{i,j}\chi_{ij}(t)\langle\Phi_{i}^{N-1}|\hat{\rho}%
(\mathbf{r})|\Phi_{j}^{N-1}\rangle, \label{eq:Q_coherences}%
\end{equation}
where $\Phi_{0}^{N}$ is the ground electronic state of the neutral, $\Phi
_{i}^{N-1}$ are the electronic states of the ion, and coefficients $\chi
_{ij}(t)$ denote the populations of electronic states when $i=j$ and the
electronic coherences when $i\neq j$.

\begin{figure}
\captionsetup[subfigure]{labelformat=empty} \subfloat{
		\includegraphics[width=0.3\textwidth]{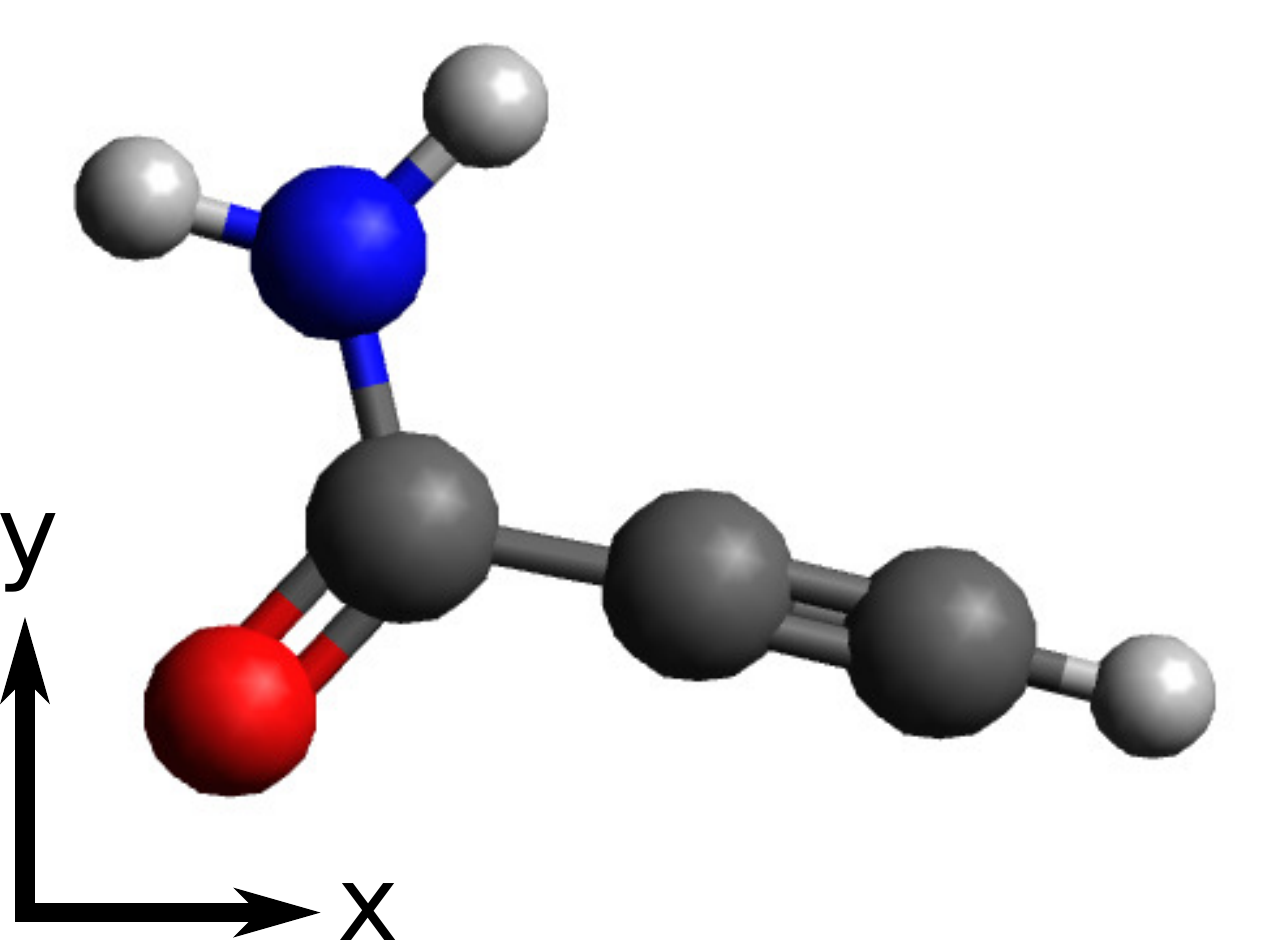}
	}

\subfloat[(a) Ionization out of HOMO, $y$ direction.]{{
		\includegraphics[width=0.45\textwidth]{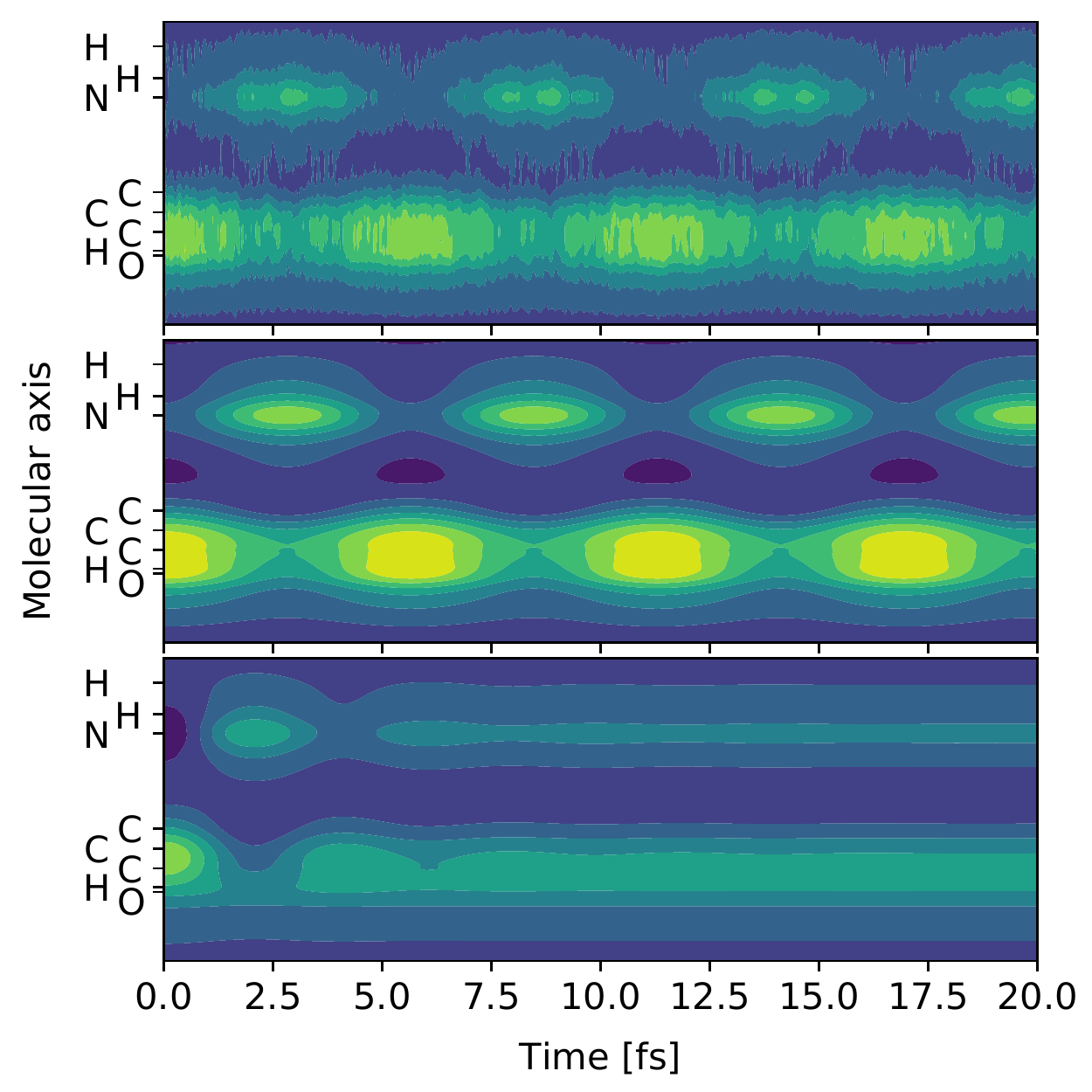}
	}} \subfloat[(b) Ionization out of HOMO--1, $x$ direction.]{{
	\includegraphics[width=0.45\textwidth]{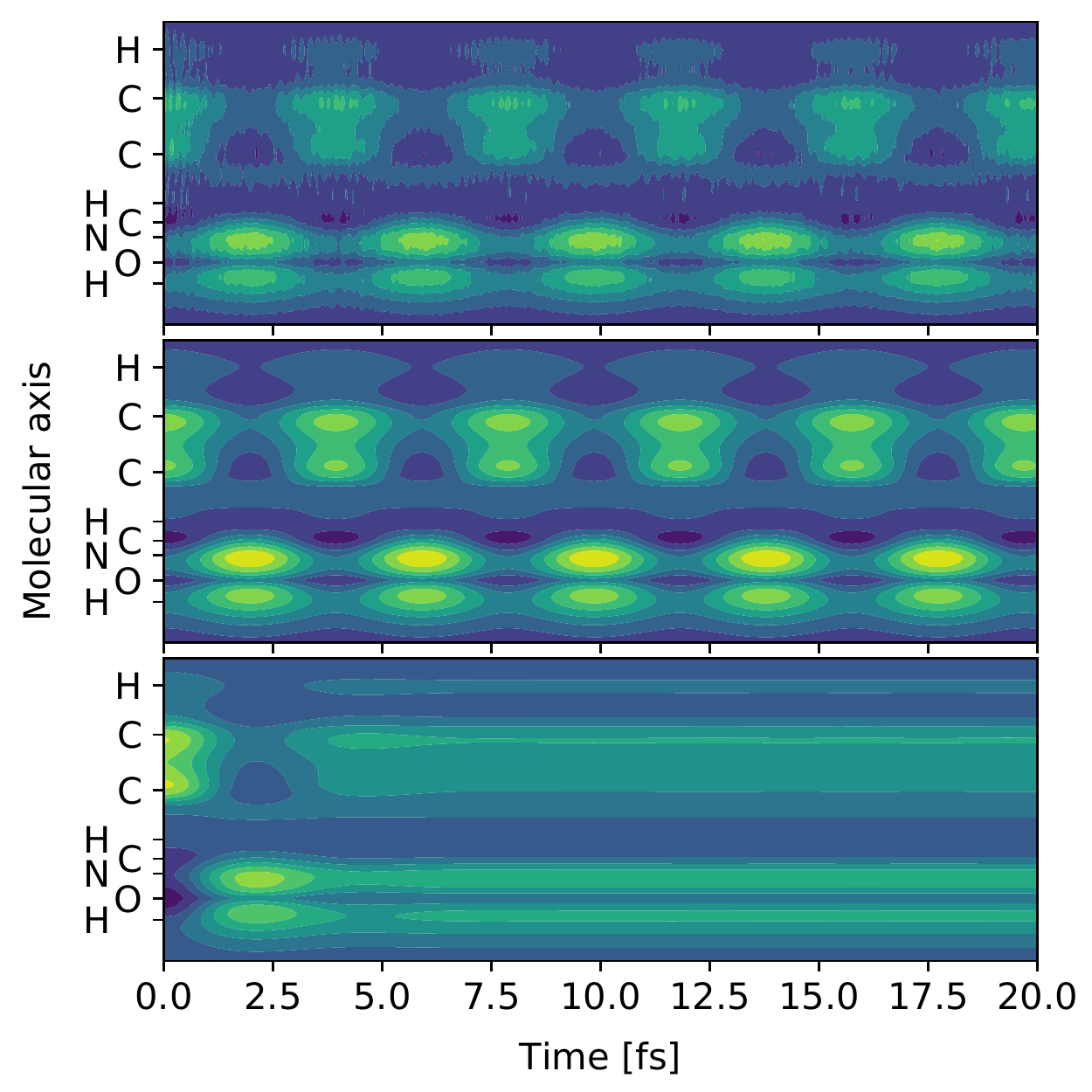}
}}
\caption{Time evolution of the hole density, Eq.~(\ref{eq:Q_coherences}), along different directions of propiolamide
molecule, after an ionization out of (a) HOMO and (b) HOMO--1. Different panels show results of electronic dynamics
calculations obtained under different assumptions. Top panels: Full electronic basis with fixed nuclei. Middle panels:
Truncated electronic basis with fixed nuclei. Bottom panels: Truncated electronic basis with moving nuclei.}
\label{fig:propiolamide_Q}
\end{figure}

The hole density computed for ionization out of HOMO and HOMO--1 of
propiolamide is shown in Fig.~\ref{fig:propiolamide_Q}. In the top panel of
the Fig.~\ref{fig:propiolamide_Q} we plot evolution of positive charge,
Eq.~(\ref{eq:Q_full}), taking into account \textit{all} electronic states of
the propiolamide in the fixed geometry framework. Middle panel illustrates
fixed geometry calculations performed by truncating full ionic subspace to the
two lower electronic states in both symmetries. A good agreement between top
and middle panels of Fig.~\ref{fig:propiolamide_Q} verifies the validity of
considered electronic dynamics in the truncated basis. Bottom panel shows
electronic dynamics in truncated basis with moving nuclei calculated by
Eq.~(\ref{eq:Q_coherences}) using electronic coherences evaluated by the TGA
scheme described in the main text. We see that in both symmetries the nuclear
motion leads to fast decoherence of electronic oscillations.

\section{Comparison of TGA with vMCG}

The performance of the vMCG method in the VC model of propiolic acid was
compared with the MCTDH benchmark and the TGA results. Both single-set ansatz
(where only one set of Gaussian functions is used for all electronic states)
and multi-set ansatz (where the Gaussians on different states evolve
separately) were considered. The single-Hessian TGA approach used in the
present study can be considered as a special case of the vMCG method. In
particular, a \textit{single} Gaussian function with \textit{relaxed}
parameters is used for each involved electronic state (multi-set ansatz), the
center of each Gaussian follows \textit{classical} (i.e., non-variational)
molecular dynamics trajectory, all nonadiabatic couplings between surfaces are
neglected, and the potential is assumed to be locally harmonic with a constant
(reference) Hessian. Indeed, due to the generality of vMCG, nearly all
Gaussian-based methods can be considered extensions or special cases of it.

vMCG calculations were performed in Quantics v1.2
package~\cite{quantics,Worth:2020}. Default options in Quantics were used for
the initial positions and widths of the Gaussian basis functions: shells in
position space with the central Gaussian fitting exactly the initial state
(widths of all Gaussians were equal to the width of the initial state).

\begin{figure}
\includegraphics[width=0.9\textwidth]{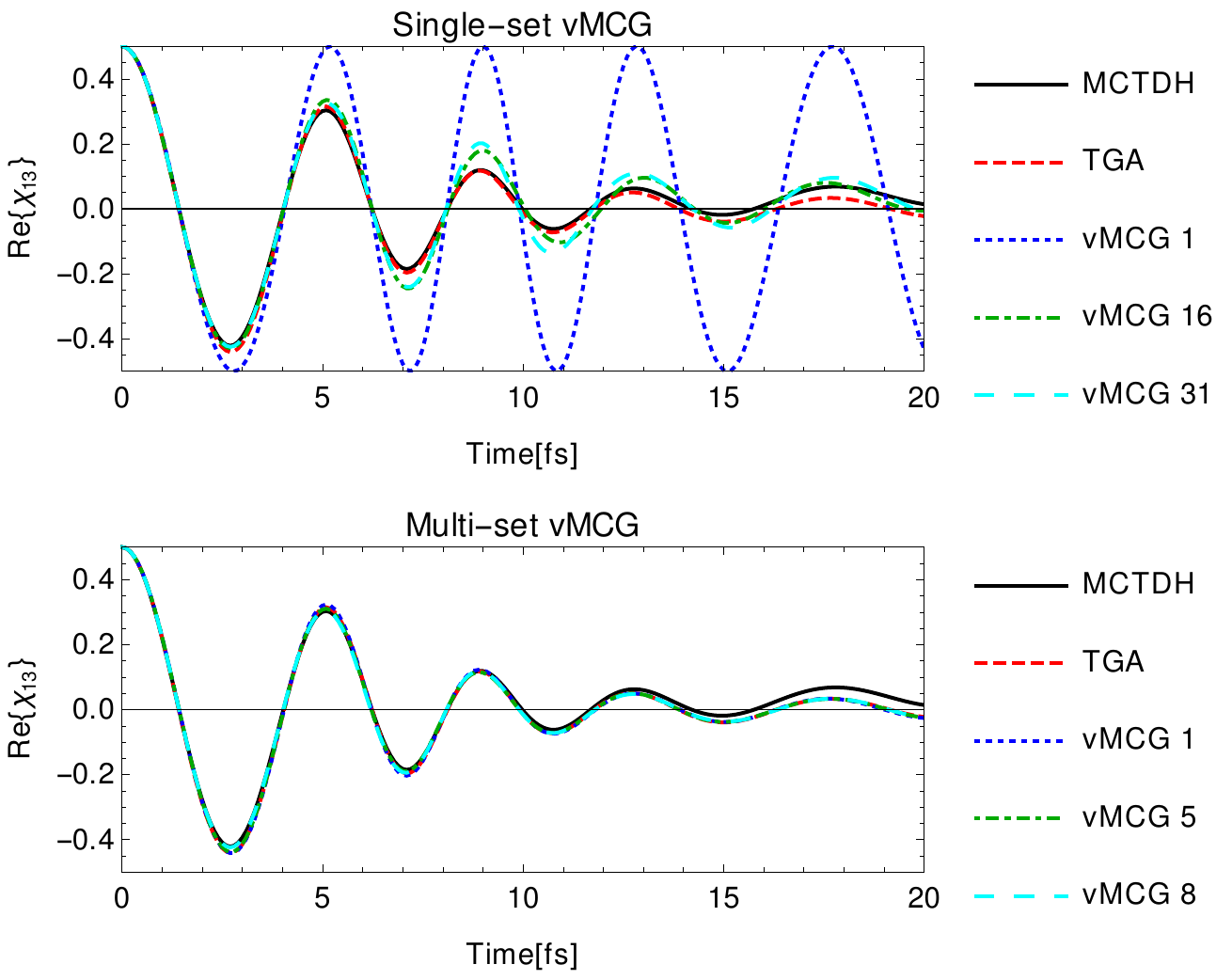}
\caption{Electronic coherence in the full-dimensional vibronic
coupling model of the propiolic acid. Comparison between the MCTDH benchmark,  the TGA result, and vMCG
calculations with variable number of frozen Gaussian basis functions.
Top panel: Single-set vMCG calculations with 1, 16, and 31 Gaussians.
Bottom panel: Multi-set vMCG calculations with 1, 5, and 8 Gaussians per state.}
\label{fig:quantics_propiolic}
\end{figure}

Figure~\ref{fig:quantics_propiolic} shows that the single-set vMCG approach
used previously in Ref.~\cite{Vacher_Malhado:2017} (Gaussian functions with
frozen parameters, the center of each Gaussian follows non-classical
variational equations of motion) cannot reproduce the exact benchmark results
even with a rather large number of basis functions, whereas the multi-set
methods perform better already with a single Gaussian function per state.
Interestingly, using more Gaussians in the multi-set vMCG method converges
only very slowly to the exact MCTDH\ non-adiabatic result. Multi-set
calculations with more than 8 Gaussians per state were failing---further
convergence of the vMCG results would require careful tuning of the initial
parameters of the Gaussian functions (see, e.g., Ref.~\cite{Jenkins_Robb:2018}%
). Remarkably, the TGA, which uses only a single classical trajectory (for
each of the two states), yields in this system better results than the
single-set version of vMCG with 31 variational trajectories and as good
results as the single-set version of vMCG with 8 variational
trajectories.\textbf{ }This is probably due to the flexible width of the
thawed Gaussian which captures more anharmonicity than a single frozen
Gaussian~\cite{Han_Guo:2018} (but which can, in multi-trajectory coupled
thawed Gaussian methods, result in numerical instabilities).

As for the cost of the calculations, the single-set approach is
computationally advantageous when a very large number of electronic states is
considered---however, there is no guarantee that the method will converge
within a feasible number of basis functions. In contrast, the multi-set
methods employ a different Gaussian basis on each involved electronic state
and, therefore, much fewer basis functions per state are needed. According to
Fig.~\ref{fig:quantics_propiolic}, one can expect that using only one Gaussian
per surface provides a good estimate of the short-time electronic coherence.
In systems with stronger anharmonic effects and larger differences between
surface curvatures, an additional degree of accuracy can be achieved by using
the TGA, which allows the Gaussian wavepacket to spread or contract, but
requires additional Hessian evaluations (one per surface in the single-Hessian
version). Finally, since the TGA neglects nonadiabatic coupling, wavepacket
propagations in different electronic states are independent and can be
performed in parallel and only those states that are populated initially have
to be considered, which makes the calculations extremely efficient.
Interestingly, the importance of nonadiabatic couplings and thus the
applicability of the TGA to the system under study can be evaluated on the fly
with an efficient semiclassical approach to estimate adiabaticity (the
so-called multiple-surface dephasing
representation~\cite{Zimmermann_Vanicek:2010,Zimmermann_Vanicek:2012}). This
method does not even require calculations of Hessians and operates only with
classical molecular dynamics trajectories.


%